\documentclass[adp,fleqn]{w-art}
\usepackage{amsmath,amssymb,w-thm,cite,times}
\usepackage{hyperref}

\usepackage[]{graphicx}

\newcommand{\ud}{\mathrm{d}}
\newcommand{\ve}{\varepsilon}

\begin{document}

\title{Three steps to accelerated expansion}

\author[O. Hrycyna]
{Orest Hrycyna\inst{1,}\footnote{E-mail:~\textsf{hrycyna@kul.lublin.pl}}}
\author[M. Szyd{\l}owski]
{Marek
Szyd{\l}owski\inst{2,3,}\footnote{E-mail:~\textsf{uoszydlo@cyf-kr.edu.pl}}}

\address[\inst{1}]{Department of Theoretical Physics, Faculty of Philosophy,
The John Paul II Catholic University of Lublin, Al. Rac{\l}awickie 14, 20-950
Lublin, Poland}
\address[\inst{2}]{Astronomical Observatory, Jagiellonian University, Orla 171,
30-244 Krak{\'o}w, Poland}
\address[\inst{3}]{Mark Kac Complex Systems Research Centre, Jagiellonian
University, Reymonta 4, 30-059 Krak{\'o}w, Poland}

\begin{abstract}
We study the dynamics of a non-minimally coupled scalar field cosmology with a potential
function. We use the framework of dynamical systems theory to investigate all evolutional
paths admissible for all initial conditions. Additionally, we assume the presence of
barotropic matter and show that the dynamics can be formulated in terms of an autonomous
dynamical system. We have found fixed points corresponding to three main stages of the
evolution of the universe, namely, radiation, matter and quintessence domination epochs.
Using the linearization of the dynamical systems in the vicinity of the critical points
we explicitly obtain formulas determining the effective equation of state parameter for
the universe at different epochs. In our approach the form of $w(z)$ parametrisation is
derived directly from the dynamical equations rather than postulated {\it a priori}.
\end{abstract}

\keywords{Modified gravity, dark energy theory, scalar field, non-minimal coupling.}
\subjclass[pacs]{04.50.Kd, 98.80.Cq, 95.36.+x}

\maketitle

There are principally two approaches in searching for the form of the equation of state
(EoS) for the current Universe in the accelerating phase of expansion. In the first
approach the form of the coefficient $w_{X}(z)$ of the equation of state for dark energy
is assumed at the very beginning, usually in the simple linear form with respect to the
redshift $z$ or the scale factor $a$
\cite{Chevallier:2000qy,Linder:2004ng}.
In the second approach the form of the EoS parameterisation is derived from the exact
dynamics of the underlying theoretical model \cite{Hrycyna:2007mq, Hrycyna:2007gd,
Szydlowski:2008zza, Kurek:2007bu, Kurek:2008qt}. In the present paper we realise this
idea for the Friedmann--Robertson--Walker (FRW) models filled with a scalar field
non-minimally coupled to gravity.

In our study of evolutional scenarios it is natural to use the framework of
dynamical system theory because it offers the possibility of investigations of
all solutions admissible for all initial conditions. We investigate fixed
points of a dynamical system and then linearise the system around them to find
exact forms of the EoS parameterisations. Therefore we derive $w(z)$
parameterisations directly from the underlying dynamics of the model.

We assume the spatially flat FRW universe filled with a non-minimally coupled scalar
field and barotropic fluid with a general equation of state parameter $w_{m}$, and the
action is
\begin{equation}
S = \frac{1}{2}\int \ud^{4}x \sqrt{-g} \left[\frac{1}{\kappa^{2}}R - \ve
\left(g^{\mu\nu}\partial_{\mu}\phi\partial_{\nu}\phi + \xi R \phi^{2}\right) -
2U(\phi) \right] + S_{m}\, ,
\end{equation}
where $\kappa^{2}=8\pi G$, $\ve = +1,-1$ corresponds to canonical and phantom scalar
field, respectively, and the metric signature is $(-,+,+,+)$. $S_{m}$ is the action for
the barotropic matter part.

We can obtain the dynamical equation for the scalar field from the variation $\delta
S/\delta \phi = 0$, and energy constraint from the variation $\delta S/\delta g=0\,$:
\begin{equation}
\ddot{\phi} + 3 H \dot{\phi} + \xi R \phi + \ve U'(\phi) =0\, , \quad
\mathcal{E}= \ve \frac{1}{2}\dot{\phi}^{2} + \ve3\xi H^{2}\phi^{2} + \ve3\xi H
(\phi^{2})\dot{} + U(\phi) + \rho_{m} - \frac{3}{\kappa^{2}}H^{2}\, .
\end{equation}
Then the conservation conditions read
\begin{equation}
\frac{3}{\kappa^{2}}H^{2} =  \rho_{\phi} + \rho_{m}\, ,\qquad
\dot{H} =  -\frac{\kappa^{2}}{2}\left[(\rho_{\phi}+p_{\phi}) +
\rho_{m}(1+w_{m})\right]\, ,
\end{equation}
where the energy density and the pressure of the scalar field are
\begin{eqnarray}
\rho_{\phi} = \ve\frac{1}{2}\dot{\phi}^{2}+U(\phi)+\ve3\xi H^{2}\phi^{2} +
\ve 3\xi H (\phi^{2})\dot{}\, ,\\
p_{\phi} = \ve\frac{1}{2}(1-4\xi)\dot{\phi}^{2} - U(\phi) + \ve\xi
H(\phi^{2})\dot{} - \ve2\xi(1-6\xi)\dot{H}\phi^{2} -
\ve3\xi(1-8\xi)H^{2}\phi^{2} + 2\xi\phi U'(\phi)\, .
\end{eqnarray}

In what follows we introduce the energy phase space variables
\begin{equation}
x\equiv \frac{\kappa \dot{\phi}}{\sqrt{6}H}, \quad
y\equiv\frac{\kappa\sqrt{U(\phi)}}{\sqrt{3}H}, \quad
z\equiv\frac{\kappa}{\sqrt{6}}\phi,
\end{equation}
which are suggested by the conservation condition
$\frac{\kappa^{2}}{3H^{2}}\rho_{\phi} + \frac{\kappa^{2}}{3H^{2}}\rho_{m} = \Omega_{\phi}
+ \Omega_{m} = 1$.

The acceleration equation can be written in the form
\begin{equation}
\dot{H} = -\frac{\kappa^{2}}{2}\left(\rho_{\rm{eff}}+p_{\rm{eff}}\right) =
-\frac{3}{2}H^{2}(1+w_{\rm{eff}}) \, ,
\end{equation}
where the effective equation of state parameter reads $\left(\lambda =
-\frac{\sqrt{6}}{\kappa}\frac{1}{U(\phi)}\frac{\ud U(\phi)} {\ud\phi} \right)$
\begin{multline}
w_{\rm{eff}}=\frac{1}{1-\ve6\xi(1-6\xi)z^{2}}\Big[ -1 +
\ve(1-6\xi)(1-w_{m})x^{2} + \ve2\xi(1-3w_{m})(x+z)^{2} + \\
 + (1+w_{m})(1-y^{2}) -\ve2\xi(1-6\xi)z^{2} - 2\xi\lambda y^{2} z\Big] \, .
 \label{eq:weff}
\end{multline}

\begin{table}
\caption{The location and eigenvalues of the critical points in twister
quintessence scenario}
\label{tab:1}
\begin{tabular}{|c|c|c|}
\hline
$w_{\rm{eff}}$ & location & eigenvalues \\
\hline
$\frac{1}{3}$ & $x_{1}^{*}=0, y_{1}^{*}=0,
(\lambda_{1}^{*})^{2}=\frac{\alpha^{2}}{\ve6\xi}$ & $l_{1}=-6\xi$,
$l_{2}=12\xi$, $l_{3}=6\xi(1-3w_{m})$ \\
$w_{m}$ & $x_{2}^{*}=0, y_{2}^{*}=0, \lambda_{2}^{*}=0$ &
$l_{1,3}=-\frac{3}{4}(1-w_{m})\Big(1\pm\sqrt{1-\frac{16}{3}\xi\frac{1-3w_{m}}{(1-w_{m})^{2}}}\Big)$,
$l_{2}=\frac{3}{2}(1+w_{m})$\\
$-1$ & $x_{3}^{*}=0, (y_{3}^{*})^{2}=1, \lambda_{3}^{*}=0 $ &
$l_{1,3}=-\frac{1}{2}\Big(3\pm\sqrt{9+\ve2\alpha-48\xi}\Big)$,
$l_{2}=-3(1+w_{m})$\\
\hline
\end{tabular}
\end{table}

The dynamical system of the model under consideration takes the form
\cite{Szydlowski:2008in, Hrycyna:2009zj}
\begin{align}
x' & = -3x -12\xi z +\ve\frac{1}{2}\lambda y^{2}\Big[1-\ve6\xi z(x+z)\Big]+
\ve 6\xi(1-6\xi)x z^{2} + \nonumber \\
& + \frac{3}{2}(x+6\xi z)\bigg[\ve(1-6\xi)(1-w_{m})x^{2} +
\ve2\xi(1-3w_{m})(x+z)^{2} + (1+w_{m})(1-y^{2})\bigg], \\
y' & = -\frac{1}{2}\lambda y\bigg\{x\Big[1-\ve6\xi(1-6\xi)z^{2}\Big] + 6\xi
y^{2}z\bigg\} - \ve12\xi(1-6\xi)y z^{2}
+ \nonumber \\
& + \frac{3}{2} y \bigg[\ve(1-6\xi)(1-w_{m})x^{2} +
\ve2\xi(1-3w_{m})(x+z)^{2} + (1+w_{m})(1-y^{2})\bigg], \\
z' & = x \Big[1-\ve6\xi(1-6\xi)z^{2}\Big], \\
\lambda' & = -\lambda^{2}(\Gamma - 1)x\Big[1-\ve6\xi(1-6\xi)z^{2}\Big],
\end{align}
where $\Gamma = \frac{U''(\phi)U(\phi)}{U'(\phi)^{2}}$ and a prime denotes
differentiation with respect to time $\tau$ defined as $ \frac{\ud}{\ud \tau} =
\Big[1-\ve6\xi(1-6\xi)z^{2}\Big] \frac{\ud}{\ud \ln{a}} $. In the rest of the paper we
will assume the function $\Gamma$ in the form
$\Gamma(\lambda)=1-\frac{\alpha}{\lambda^{2}}$. In our approach the exact
formulas for $w(z)$ parameterisations can be derived directly
from the linearised solutions of the dynamical system in the vicinity of the
critical
points representing different epochs in the evolution of the universe
(Table~\ref{tab:1}). In the special case of conformally coupled ($\xi=1/6$)
canonical
($\ve=1$) scalar field and dust matter ($w_{m}=0$), they are as follows:

\noindent 1) At the radiation domination epoch,
\begin{multline}
w_{\rm{eff}}^{R}(z)|_{\xi=\frac{1}{6}} = \frac{1}{3} +
\frac{2\lambda_{1}^{*}}{3\alpha}B_{1} \left(\frac{1+z}{1+z_{R}}\right)^{-1} +
\left(\frac{1}{3}B_{1}^{2}-\frac{\alpha}{6}A_{1}^{2}\left(y_{1}^{(i)}\right)^{2}\right)
\left(\frac{1+z}{1+z_{R}}\right)^{-2} + \\ +
\frac{1}{3}\lambda_{1}^{*}A_{1}\left(y_{1}^{(i)}\right)^{2}
\left(\frac{1+z}{1+z_{R}}\right)^{-3}
- \left(1+\frac{\alpha}{3}-\frac{\alpha}{6}A_{1}B_{1}\right)\left(y_{1}^{(i)}\right)^{2}
\left(\frac{1+z}{1+z_{R}}\right)^{-4} - \\ -
\frac{1}{3}\lambda_{1}^{*}B_{1}\left(y_{1}^{(i)}\right)^{2}
\left(\frac{1+z}{1+z_{R}}\right)^{-5}
- \frac{\alpha}{12}B_{1}^{2}\left(y_{1}^{(i)}\right)^{2}
\left(\frac{1+z}{1+z_{R}}\right)^{-6} \, ,
\end{multline}
where
$A_{1}=x_{1}^{(i)}-\frac{1}{\alpha}\left(\lambda_{1}^{(i)}-\lambda_{1}^{*}\right)$,
$B_{1}=x_{1}^{(i)}+\frac{1}{\alpha}\left(\lambda_{1}^{(i)}-\lambda_{1}^{*}\right)$,
$\left(\lambda_{1}^{*}\right)^{2}=\alpha^{2}$ and $z_{R}$ is the redshift of
the radiation domination epoch.

\noindent 2) At the matter domination epoch,
\begin{multline}
w_{\rm{eff}}^{M}(z)|_{\xi=\frac{1}{6}} =
-\alpha\frac{4}{3}A_{2}^{2}\left(y_{2}^{(i)}\right)^{2}
\left(\frac{1+z}{1+z_{M}}\right)^{-1} +
\alpha\frac{8}{3}A_{2}B_{2} \left(y_{2}^{(i)}\right)^{2}
\left(\frac{1+z}{1+z_{M}}\right)^{-3/2} - \\
-\alpha\frac{4}{3}B_{2}^{2}\left(y_{2}^{(i)}\right)^{2}
\left(\frac{1+z}{1+z_{M}}\right)^{-2} -
\left(y_{2}^{(i)}\right)^{2} \left(\frac{1+z}{1+z_{M}}\right)^{-3} \, ,
\end{multline}
where $A_{2}=x_{2}^{(i)}+\frac{\lambda_{2}^{(i)}}{2\alpha}$,
$B_{2}=x_{2}^{(i)}+\frac{\lambda_{2}^{(i)}}{\alpha}$ and $z_{M}$ is the redshift
of the matter domination epoch.

\noindent 3) In the de Sitter state, we have two parameterisations whose type depends on
the value of the parameter $\alpha$ characterising the shape
of the function $\Gamma$. \\
--- For $\alpha>-1/2$, we have a linear approach to the
de Sitter state
\begin{multline}
w_{\rm{eff}}^{Q}(z)|_{\xi=\frac{1}{6}} = -1 +
\frac{1}{6\Delta}\left(1-\alpha+\sqrt{\Delta}\right)A_{3}^{2}
(1+z)^{3+\sqrt{\Delta}} + \\ +
\left(\frac{\alpha}{\Delta}A_{3}B_{3} - 2
y_{3}^{*}\left(y_{3}^{(i)}-y_{3}^{*}\right) \right) (1+z)^{3} +
\frac{1}{6\Delta}
\left(1-\alpha-\sqrt{\Delta}\right)B_{3}^{2} (1+z)^{3-\sqrt{\Delta}} - \\ -
\frac{2\alpha}{3\Delta} A_{3}^{2}
y_{3}^{*}\left(y_{3}^{(i)}-y_{3}^{*}\right) (1+z)^{6+\sqrt{\Delta}} +
\left(\frac{4\alpha}{3\Delta}A_{3}B_{3}y_{3}^{*}-y_{3}^{(i)}+
y_{3}^{*}\right) \left(y_{3}^{(i)}-y_{3}^{*}\right)(1+z)^{6} - \\ -
\frac{2\alpha}{3\Delta}B_{3}^{2} y_{3}^{*}\left(y_{3}^{(i)}-y_{3}^{*}\right)
(1+z)^{6-\sqrt{\Delta}} - \\ -
\frac{\alpha}{3\Delta}\left(y_{3}^{(i)}-y_{3}^{*}\right)^{2}
\left(A_{3}(1+z)^{\frac{\sqrt{\Delta}}{2}}+
B_{3}(1+z)^{-\frac{\sqrt{\Delta}}{2}}\right)^{2} (1+z)^{9}
\end{multline}
where $A_{3}=x_{3}^{(i)}+\frac{1}{2\alpha}(3-\sqrt{\Delta})\lambda_{3}^{(i)}$,
$B_{3}=x_{3}^{(i)}+\frac{1}{2\alpha}(3+\sqrt{\Delta})\lambda_{3}^{(i)}$,  and
$\Delta=1+2\alpha >0$. \\
--- For $\alpha<-1/2$, we have a damped oscillatory approach to the de
Sitter state
\begin{multline}
w_{\rm{eff}}^{Q}(z)|_{\xi=\frac{1}{6}} =
- 1 - 2y_{3}^{*}\left(y_{3}^{(i)}-y_{3}^{*}\right) (1+z)^{3} + \\ +
\frac{1}{3|\Delta|}
\left(\left(C_{3}+\frac{|\Delta|\lambda_{3}^{(i)}}{2\alpha}\right)^{2} - \alpha
4 C_{3}^{2}\right)(1+z)^{3}\sin^{2}{\left(\frac{\sqrt{|\Delta|}}{2}\ln{(1+z)}\right)}
+ \\ +
\frac{1}{3\sqrt{|\Delta|}}
\left(\left(C_{3}+\frac{|\Delta|\lambda_{3}^{(i)}}{2\alpha}\right)
\left(x_{3}^{(i)}+\frac{\lambda_{3}^{(i)}}{\alpha}\right) + 2
C_{3}\frac{\lambda_{3}^{(i)}}{\alpha}\right)(1+z)^{3}
\sin{\left(\sqrt{|\Delta|}\ln{(1+z)}\right)} + \\ +
\frac{1}{3}\left(\left(x_{3}^{(i)}+\frac{\lambda_{3}^{(i)}}{\alpha}\right)^{2}
-\alpha\left(\frac{\lambda_{3}^{(i)}}{\alpha}\right)^{2}\right) (1+z)^{3}
\cos^{2}{\left(\frac{\sqrt{|\Delta|}}{2}\ln{(1+z)}\right)} - \\ -
\frac{4\alpha}{3|\Delta|}
\left(2y_{3}^{*}\left(y_{3}^{(i)}-y_{3}^{*}\right) +
\left(y_{3}^{(i)}-y_{3}^{*}\right)^{2}(1+z)^{3}\right) \\
\left(C_{3}\sin{\left(\frac{\sqrt{|\Delta|}}{2}\ln{(1+z)}\right)} -
\frac{\sqrt{|\Delta|}\lambda_{3}^{(i)}}{2\alpha}
\cos{\left(\frac{\sqrt{|\Delta|}}{2}\ln{(1+z)}\right)}\right)^{2} (1+z)^{6} \, ,
\end{multline}
where $C_{3}=x_{3}^{(i)}+\frac{3\lambda_{3}^{(i)}}{2\alpha}$,
$|\Delta|=-1-2\alpha>0$ and $z_{Q}=0$ is the redshift of the present time.

These parameterisations depend on three values $x_{n}^{(i)}$, $y_{n}^{(i)}$,
$\lambda_{n}^{(i)}$ of the initial conditions for the linearised solutions at different
epochs and the parameter $\alpha$ describing the shape of the function $\Gamma(\lambda)$.
The common parameter $\alpha$ is present in all parameterisations and can be estimated
from the observational data.

\begin{figure}
\sidecaption
\includegraphics[scale=0.43]{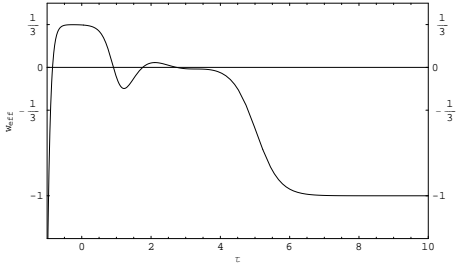}
\caption{The evolution of $w_{\rm{eff}}$ given by the relation
(\ref{eq:weff}) for the non-minimally
coupled canonical scalar field $\ve=+1$ and the positive coupling constant
$\xi$. The existence of a short
time interval during which $w_{\rm{eff}}\simeq\frac{1}{3}$ is the effect
of the nonzero coupling constant $\xi$.}
\label{fig:2}
\end{figure}

In this short note, we presented the possibility of extracting the equation of state
parameterisations of dynamical dark energy directly from the dynamics of the underlying
theoretical model. It is interesting that the obtained formulas for $w(z)$ are all that
is needed for the realistic cosmological model, i.e., the radiation epoch required by the
nucleosynthesis, matter domination phase and final acceleration epoch (see
Fig.~\ref{fig:2}).

\providecommand{\WileyBibTextsc}{}
\let\textsc\WileyBibTextsc
\providecommand{\othercit}{}
\providecommand{\jr}[1]{#1}
\providecommand{\etal}{~et~al.}


\begin{thebibliography}{[1]}

\bibitem{Chevallier:2000qy}
 \textsc{M.~Chevallier} and  \textsc{D.~Polarski},
 \jr{Int. J. Mod. Phys.} \textbf{D10}, 213--224 (2001),
\href{http://arxiv.org/abs/gr-qc/0009008}{arXiv:gr-qc/0009008}.


\bibitem{Linder:2004ng}
 \textsc{E.\,V. Linder},
 \jr{Phys. Rev.} \textbf{D70}, 023511 (2004),
\href{http://arxiv.org/abs/astro-ph/0402503}{arXiv:astro-ph/0402503}.


\bibitem{Hrycyna:2007mq}
 \textsc{O.~Hrycyna} and  \textsc{M.~Szydlowski},
 \jr{Phys. Lett.} \textbf{B651}, 8--14 (2007),
\href{http://arxiv.org/abs/0704.1651}{arXiv:0704.1651}.


\bibitem{Hrycyna:2007gd}
 \textsc{O.~Hrycyna} and  \textsc{M.~Szydlowski},
 \jr{Phys. Rev.} \textbf{D76}, 123510 (2007),
\href{http://arxiv.org/abs/0707.4471}{arXiv:0707.4471}.


\bibitem{Szydlowski:2008zza}
 \textsc{M.~Szydlowski},  \textsc{O.~Hrycyna},  and  \textsc{A.~Kurek},
 \jr{Phys. Rev.} \textbf{D77}, 027302 (2008),
\href{http://arxiv.org/abs/0710.0366}{arXiv:0710.0366}.


\bibitem{Kurek:2007bu}
 \textsc{A.~Kurek},  \textsc{O.~Hrycyna},  and  \textsc{M.~Szydlowski},
 \jr{Phys. Lett.} \textbf{B659}, 14--25 (2008),
\href{http://arxiv.org/abs/0707.0292}{arXiv:0707.0292}.


\bibitem{Kurek:2008qt}
 \textsc{A.~Kurek},  \textsc{O.~Hrycyna},  and  \textsc{M.~Szydlowski},
  {From model dynamics to oscillating dark energy parametrisation}, 
\href{http://arxiv.org/abs/0805.4005}{arXiv:0805.4005}.


\bibitem{Szydlowski:2008in}
 \textsc{M.~Szydlowski} and  \textsc{O.~Hrycyna},
 \jr{JCAP} \textbf{01}, 039 (2009),
\href{http://arxiv.org/abs/0811.1493}{arXiv:0811.1493}.


\bibitem{Hrycyna:2009zj}
 \textsc{O.~Hrycyna} and  \textsc{M.~Szydlowski}, Twister quintessence
  scenario,
\href{http://arxiv.org/abs/0906.0335}{arXiv:0906.0335}.


\end{thebibliography}
\end{document}